\documentstyle[12pt,epsfig]{article}
\begin{document}
\title{Cosmological models with one extra dimension.}
\author{A.Yu.Neronov\\ 
 Theoretische Physik, \\
Universit\"at M\"unchen, D-80333, 
M\"unchen, Germany}
\maketitle
\footnote{e-mail: neronov@theorie.physik.uni-muenchen.de}
\begin{abstract}
We consider cosmological models in which  a
homogeneous isotropic universe is embedded as a 3+1 dimensional 
surface  into a 4+1 dimensional 
manifold. The size of the extra dimension depends on time.  It 
is small compared to the size of the universe only if  
the energy of gravitational self-interaction of the universe 
through the  compact extra dimension dominates over all other kinds of energy. 
The self-interaction energy gives the main contribution
into the Friedmann equation, which governs the dynamics of the scale factor
of the universe.
\end{abstract}

The possibility that space-time has more than three spatial dimensions
arose for the first time in the Kaluza-Klein approach to unification of
gravity with electromagnetism \cite{kaluza}. This idea is implemented now  
in different models of quantum field theory and string theory. It is natural 
to ask whether the presence of additional space-time dimensions 
could be tested experimentally. Of course, if the size 
$L$ of the additional compact dimensions is  small compared to all 
experimentally attainable scales (for example,
of order of the Planck length $L\sim L_{pl}$) 
it would be extremely difficult to develop an experiment in which this size 
could be determined directly.  Recently 
N.Arkani-Hamed {\it et al.} \cite{dimo} proposed a framework for solving the 
hierarchy problem in which the space-time manifold contains ``large'' 
extra dimensions with the size $L\gg L_{pl}$. Within this 
framework the conventional $(3+1)$ dimensional space-time $M_{3+1}$ is 
embedded as a surface into a higher dimensional manifold 
${\cal M}_{(3+n)+1}$. All the matter fields are confined to live on the 
surface $M_{3+1}$ while the gravitational field can propagate in 
${\cal M}_{(3+n)+1}$. According to \cite{dimo}, the conventional 
four-dimensional gravitational constant $G_4$ is a derived quantity, it
is related to the fundamental $4+n$ dimensional gravitational constant
$G_{4+n}$ through the size $L$ of the extra dimensions
\begin{equation}
\label{eq:gg}
G_4\sim\frac{G_{4+n}}{L^n}
\end{equation}  
Therefore, although the  
four dimensional Planck mass $M_{pl}=\sqrt{\hbar c/G_4}$ is of order of
$10^{19}$ GeV, the fundamental $4+n$ dimensional Planck mass 
$\tilde M_{pl} =\left( \hbar^{1+n} c^{1-n}/G_{4+n}\right)^{1/(2+n)}$
could be of  order of TeV. This opens the  possibility to detect the presence 
of extra dimensions experimentally in collider experiments \cite{collider}.

Apart from the direct consequences for particle physics, the presence of
extra dimensions would have a set of  testable predictions in cosmology. 
If we want to check whether the possibility that space-time has compact
extra dimensions is allowed from the cosmological point of view, we 
need first to find the solutions of Einstein equations which describe 
a homogeneous isotropic universe in this framework. In this Letter we 
concentrate our attention on cosmological models with one extra dimension.
Although the case of just one 
extra dimension is not favored in the original argumentation of \cite{dimo},
it deserves attention due to its simplicity. Besides, if the number of large
extra
dimensions if bigger than one, it is not necessary that all of them have 
equal size $L_1,...,L_n=L$. If the sizes are arranged hierarchically, we
could have $L_1\gg L_i,\ i>1$ and in this case the problem with just one 
extra dimension could serve as a first approximation to the complete problem.  

Cosmological models with extra dimensions can differ significantly from 
four-dimensional ones. When the universe surface $M_{3+1}$ is embedded into 
a higher dimensional manifold ${\cal M}_{(3+n)+1}$ it can self-interact
through the $(3+n)$ dimensional volume. 
This self-interaction could affect somehow
on the internal geometry of the surface $M_{3+1}$, that is  the dynamics of 
the scale factor of the universe. Besides, this interaction 
 becomes stronger if  the size of  the extra dimensions decreases. 
Thus, even if the size of the extra 
dimensions is very small, a cosmology with extra dimensions could be very
different from the 
standard four-dimensional one. 
There were recently some attempts to develop cosmological models with 
one extra dimension \cite{binetry,csaki}. In these models the effect of 
self-interaction of the universe through the 
additional compact dimension could not
be traced, because the authors introduce a ``hidden brane'' apart
from the ``visible'' one into ${\cal M}_{4+1}$. Besides, the Ansatz used for 
the $(4+1)$ dimensional metric implies that the size of the 
extra dimension is kept
fixed, rather than determined from Einstein equations. The authors 
of \cite{binetry,csaki} point out some problems with the realization of the  
standard scenario of primordial nucleosynthesis in their cosmological models
with one extra dimension.

In what follows we find solutions of the $(4+1)$ dimensional 
Einstein equations with 
just one brane $M_{3+1}$ which could gravitationally self-interact when
it is embedded into ${\cal M}_{4+1}$. The size of the extra dimension
is defined dynamically from Einstein equations. The condition that it must 
be much smaller than the size of the (visible part of the) universe implies
that the energy of self-interaction of the universe through the additional 
dimension dominates over all other kinds of energy in the universe.  

Let us suppose that a homogeneous isotropic $(3+1)$ dimensional 
universe with metric
\begin{equation}
\label{eq:induced}
dl^2=d\tau^2-R^2(\tau)\left( d\chi^2+f(\chi)\left( d\theta^2+
\sin^2\theta d\phi^2\right)\right)
\end{equation}
\begin{equation}
f(\chi)=\left\{
\begin{array}{cl}
\sin^2\chi & \mbox{for a closed universe}\\
\chi^2 &\mbox{for a flat universe}\\
\sinh^2\chi &\mbox{for an open universe}
\end{array}
\right.
\end{equation}
is embedded into a $(4+1)$ dimensional manifold 
$M_{3+1}\subset {\cal M}_{4+1}$. If the $(4+1)$ dimensional metric
in the  coordinates $(t,r,\chi,\theta,\phi)$ has the form 
\begin{equation}
\label{eq:metric1}
ds^2=e^{\nu(t,r)} dt^2- e^{\lambda(t,r)}dr^2-r^2
\left( d\chi^2+f(\chi)\left(
d\theta^2+\sin^2\theta d\phi^2\right)\right)
\end{equation} 
then the equation of embedding  is
\begin{equation}
\label{eq:surface}
r=R(t)
\end{equation} 
for some function $R(t)$.

The functions $\lambda(t,r),\ \nu(t,r)$ which define the 
metric (\ref{eq:metric1}) 
must be found from Einstein equations in ${\cal M}_{4+1}$. Suppose that 
all the matter is concentrated on the surface $M_{3+1}$, and the 
stress energy tensor has the form 
\begin{equation}
{\cal T}_\mu^\nu=T_\mu^\nu\delta (r-R(t)).
\end{equation}
Its projection on the surface $M_{3+1}$ is
\begin{equation}
\label{stress}
 T_i^j=\mbox{diag}(\rho, -p,-p,-p) 
\end{equation}
(indices $i,j$ run through the coordinates on $M_{3+1}$ while indices $\mu,\nu$
run over the coordinates in ${\cal M}_{3+1}$). 
Outside the surface $M_{3+1}$, the metric (\ref{eq:metric1}) is a
solution of the 
vacuum Einstein equations in $(4+1)$ dimensions. As it is shown in 
the Appendix, the functions
$\lambda,\nu$ are given by 
\begin{equation}
\label{eq:solution}
e^\nu=e^{-\lambda}=k-\frac{\mu}{r^2},
\end{equation}
where $k=+1,0,$ or $-1$ for a closed, flat and open universe, respectively.
$\mu$ is a free parameter.
The behavior (\ref{eq:solution}) of the  $(4+1)$ dimensional metric is 
easy to understand. For example, in the case of a closed universe one 
immediately recognizes in 
(\ref{eq:solution}) the generalization of the 
Schwarzschild solution to the case
of $(4+1)$ space-time dimensions \cite{perry}. 
Indeed, the metric (\ref{eq:metric1}) is  
spherically symmetric in this case and  (\ref{eq:solution}) 
describes a 
$(4+1)$ dimensional spherically-symmetric black hole. The gravitational 
mass $M$ of the black hole is related to the parameter $\mu$ as
\begin{equation}
\label{eq:grav}
\mu =\frac{8 G_5}{9 \pi}M,
\end{equation}
where $G_5$ is the gravitational constant in $(4+1)$ dimensions.
The horizon of the black hole is  situated at
\begin{equation}
\label{eq:sch}
r_g=\sqrt{\mu}.
\end{equation} 

The Einstein equations on the surface $M_{3+1}\subset {\cal M}_{4+1}$ 
can be written in the form \cite{berezin}
\begin{eqnarray}
[K_i^j]-\delta_i^j[K^l_l]=8\pi G_5 T_i^j\nonumber\\
\label{eq:einstein}
T_{i;j}^j=0\\
{K_j^i}T_i^j=0,\nonumber
\end{eqnarray}
where $K_{ij}$ is the 
extrinsic curvature tensor of $M_{3+1}$. Here the semicolon
denotes the covariant derivative with respect to the induced metric 
(\ref{eq:induced}) and the square brackets $[{\cal A}]$
denote the jump  of a function ${\cal A}(t,r,\chi,\theta,\phi)$ across the 
surface.

For the surface (\ref{eq:surface}) we have $K_2^2=K_3^3=K_4^4$ and
the stress-energy tensor (\ref{stress})  
also has a simple form. Thus, the system
(\ref{eq:einstein}) can be transformed into a more simple system
of equations
\begin{eqnarray}
\label{eq:main}
[K_2^2]=\frac{8}{3}\pi G_5 \rho\nonumber\\
\dot\rho +3\frac{\dot R}{R}(\rho+p)=0,
\end{eqnarray}
where dot denotes differentiation with respect to  ``cosmological'' 
time $d/d\tau$
(\ref{eq:induced}).
The $K_{22}$ component of the extrinsic curvature tensor for the 
surface (\ref{eq:surface}) is \cite{berezin,prd57}
\begin{equation}
\label{k22}
K_{22}=-\frac{\sigma}{R}\sqrt{\dot R^2+e^{-\lambda}}=-\frac{\sigma}{R}
\sqrt{\dot R^2+k-\frac{\mu}{R^2}}
\end{equation}
where  $\sigma=+1$ if the transversal spatial coordinate 
grows in the direction of the surface normal  and 
$\sigma =-1$ if it decreases in this direction 
(see \cite{berezin,prd57} for a 
stricter definition). 
Equations (\ref{eq:main}), (\ref{k22}) 
govern the dynamics of the scale factor of a universe which is embedded as a 
surface in a $(4+1)$ dimensional space-time.

\begin{figure}[!h]
  \begin{center}
    \epsfig{file=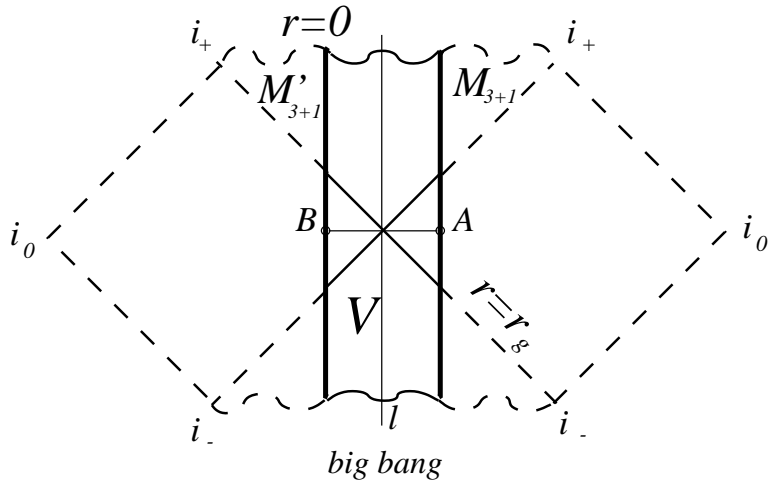}
  \end{center}
\caption{\it The structure of $(4+1)$  dimensional space-time with a
closed universe.}
\label{fig:brane1}
\end{figure}

Up to now we have discussed only the local structure of the solutions of 
Einstein equations
and payed no attention to their global properties. 
Now our task will be to find solutions with a compact extra dimension.
Let us first consider the case of a closed universe.
In this case the metric outside $M_{3+1}$ is the 
Schwarzschild metric. Therefore, 
${\cal M}_{4+1}$ is a part of the $(4+1)$ dimensional Schwarzschild manifold.
The global structure of the Schwarzschild space-time in $(4+1)$ dimensions is 
essentially the same as in $(3+1)$ dimensions. A conformal diagram for this 
space-time is presented on 
Fig. \ref{fig:brane1}. We have two asymptotically flat regions with spatial 
infinities $i_0$ and two space-like singularities (lines $i_-i_-$ and 
$i_+i_+$) in the absolute past and the absolute future. 
The surface $M_{3+1}$ is represented 
by a line which starts at the 
past singularity, then expands to a maximal radius
and then contracts to the future singularity.

A typical space-like section of this space-time has  
wormhole geometry: the coordinate $r$ changes from $\infty$ at the left 
$i_0$ to a
minimal value $r_{min}$ and then back to $\infty$ at the right $i_0$ 
along a space-like 
section. But  $r$ parameterizes the ``extra'' dimension of 
``our'' space-time $M_{3+1}$. Thus, if we want this extra dimension to 
be compact, we can not allow $r$ to change unboundly. In order to 
``compactify'' the extra dimension let us implement the following procedure.
First, we introduce another surface $M'_{3+1}$ on the other 
side of the wormhole, placed symmetrically to the surface $M_{3+1}$ with 
respect to the mouth of 
the 
wormhole (see Fig. \ref{fig:brane1}). Next, we cut the Schwarzschild manifold 
along the surfaces $M_{3+1}$ and $M'_{3+1}$ so that only the region $V$ of 
Fig. \ref{fig:brane1}  is left
after this step. The induced metrics on $M_{3+1}$ and $M'_{3+1}$ are the same
and we can  identify  $M_{3+1}\equiv M'_{3+1}$. In the obtained 
$(4+1)$ dimensional space-time ${\cal M}_{4+1}$
the coordinate $r$ changes along a spatial section from the maximal value
$r_{max}=R(\tau)$ at $M_{3+1}$ to   $r=r_{min}$ and then back to the 
maximum.
Thus, the extra dimension is compact.

\begin{figure}[!h]
  \begin{center}
    \epsfig{file=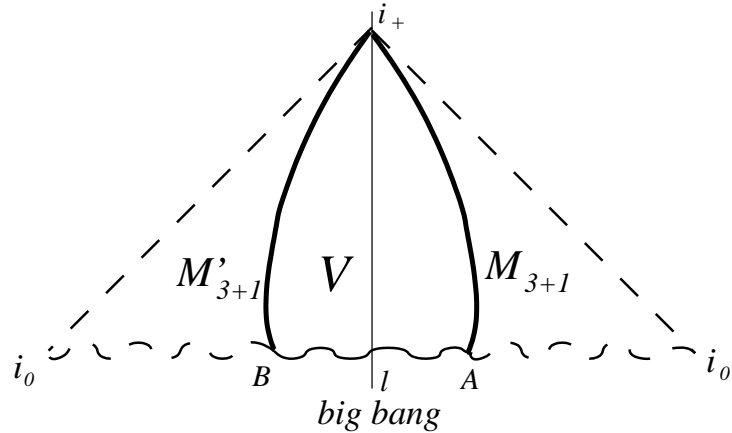}
  \end{center}
\caption{\it The structure of $(4+1)$  dimensional space-time with a
flat or open universe.}
\label{fig:brane3}
\end{figure}

The same procedure can be implemented in the case of 
open and flat universes. The global structure of the $(4+1)$ dimensional
space-time with the metric (\ref{eq:metric1}), (\ref{eq:solution}) in the cases
$k=0,-1$ and $\mu>0$ is shown on Fig. \ref{fig:brane3}. There exists only 
the past singularity $i_0i_0$. For these values of the 
parameters the coordinate $r$ is time-like and the 
coordinate $t$ parameterizes the  
``extra''dimension. The space-like section also has  wormhole geometry, 
because $t$ can change from $-\infty$ to $+\infty$ along a space-like 
section. Again, we can place the two identical surfaces 
$M_{3+1}$ and $M'_{3+1}$ at the different
sides of the 
wormhole, cut the asymptotic regions and glue the surfaces together
in order to get a compact extra dimension. The procedure is shown on Fig.
\ref{fig:brane3}.

The dynamics of the surface $M_{3+1}$  in ${\cal M}_{4+1}$
is governed by the set of equations 
(\ref{eq:main}). The space-time metric in the volume $V$ is specified
by the parameter $\mu$ which in the case of a closed universe is related to 
the mass $M$ (\ref{eq:grav}) of the 
black hole. In order to write down the final
form of equations (\ref{eq:main}) we need also to determine the 
quantity $\sigma$ in (\ref{k22}). From the definition of $\sigma$ 
we see that $\sigma =+1$ at the surface $M_{3+1}$  and 
$\sigma=-1$  at $M'_{3+1}$. 
The jump of the $K_{22}$ component of the extrinsic curvature on the 
surface $M_{3+1}=M'_{3+1}\subset {\cal M}_{4+1}$ is thus
\begin{equation}
\left[K_2^2\right]=
\frac{2}{R}\sqrt{\dot R^2+k-\frac{\mu}{R^2}}=\frac{8}{3}\pi G_5 \rho.
\end{equation}
Squaring the last equation we get an equation which is analogous to
the  conventional Friedmann equation of the $(3+1)$ dimensional cosmology 
\begin{equation}
\label{eq:adot}
\frac{\dot R^2}{R^2}+\frac{k}{R^2}=\frac{16\pi^2G_5^2 }{9} \rho^2
+\frac{\mu}{R^4}.
\end{equation}
The essential difference of (\ref{eq:adot}) from the usual Friedmann 
equation is that the energy density $\rho$ enters (\ref{eq:adot}) 
quadratically. There is an additional contribution
into the r.h.s. of the Friedmann equation in our model 
due to the self-interaction of the brane universe $M_{3+1}$  
through the compact extra
dimension. This contribution  is similar to the 
contribution which would come from   
radiation  (with equation of state $p=\rho/3$) with energy density 
\begin{equation}
\label{eq:effective}
\rho_{s}=\frac{3\mu}{8\pi G_4R^4}
\end{equation}
($G_4$ is the four-dimensional gravitational constant).

The essential requirement imposed on any cosmological model with extra 
dimensions is that the size $L$ of extra dimensions must be much smaller than 
the the parameter  $R(\tau)$ 
\begin{equation}
\label{size}
L\ll R
\end{equation}
(in the case of a closed universe
$R(\tau)$ is the size of the universe, while in the case of flat universe
we must modify this requirement and speak about the size of the visible part 
of the universe). Let us estimate the values of the parameters of our model, 
when (\ref{size}) is satisfied. In a matter dominated closed universe 
we have $p =0, k=1$. In this case
\begin{equation}
\label{eq:total}
\rho=\frac{3}{4\pi}\frac{m}{R^3},
\end{equation}
where $m$ is the total mass of the universe. Taking 
$\dot R=0$ in (\ref{eq:adot}) we find that 
the maximal expansion of the universe is  
\begin{equation}
\label{eq:amax}
R_{max}^2=\frac{1}{2}\left( \mu+\sqrt{\mu^2+4 G_5^2 m^2}
\right).
\end{equation}
The size of the 
extra dimension at the moment of maximal expansion  can be 
defined as the distance between the 
points $A$ and $B$ of Fig. \ref{fig:brane1}
along the spatial section $t=0$ of the Schwarzschild space-time
\begin{equation}
\label{eq:max}
L_{max} =\int\limits_A^{B}\frac{dr}{\sqrt{1-\mu/r^2}} =2\sqrt{R_{max}^2-\mu}.
\end{equation}
We see that the size of the extra dimension is small compared to the size of 
the universe if 
\begin{equation}
R_{max}\approx r_g=\sqrt{\mu},
\end{equation}
i.e. if the maximal scale factor is close to the 
gravitational radius of the $(4+1)$ dimensional black hole. 
From (\ref{eq:amax}) we find that
this is true when 
\begin{equation}
\label{eq:small}
M\gg m 
\end{equation}
that is the gravitational mass (\ref{eq:grav}) of the universe must be much 
greater than its bare mass  (\ref{eq:total}).

But when (\ref{eq:small}) is true, the second term in the r.h.s.
of Friedmann equation (\ref{eq:adot}) dominates over the first one. 
Neglecting the term proportional to $\rho^2$ in zero approximation
we get the following law of expansion 
\begin{equation}
\label{eq:zero}
R_l(\tau)=\sqrt{2\sqrt{\mu}\tau -\tau^2}.
\end{equation} 
We have added the subscript $l$ to  $R(\tau)$ because the dependence
(\ref{eq:zero}) is exactly the dependence of the radius on proper time along 
the line $l$ (the mouth of the wormhole) of Fig. \ref{fig:brane1}. In zero 
approximation the trajectories of both $M_{3+1}$ and $M'_{3+1}$ follow 
the line $l$,  and therefore the size of the extra dimension is equal to zero.
 
The size of the extra dimension is not constant in our model. 
When it is small, we can implement the following 
procedure in order to define $L(\tau)$. Let us introduce a locally 
Minkowsky reference frame which is comoving to the brane $M_{3+1}$ at some 
moment of time $\tau_0$. In this frame we can use 
the usual nonrelativistic definition of spatial distances (and therefore of 
the ``size'' of the extra dimension) in the vicinity of 
the brane. Since both $M_{3+1}$ and $M'_{3+1}$ move close to the line $l$  
of Fig. \ref{fig:brane1}, 
we can use a locally Minkowsky frame comoving to $l$ instead of the 
frame comoving to $M_{3+1}$.

In order to find a comoving frame at a given point $\tau_0$ of $l$ we first
introduce the analog of the Kruskal coordinate system 
in the white hole region of the $(4+1)$ 
dimensional Schwarzschild space-time
\begin{equation}
\label{eq:kruskal}
\left\{
\begin{array}{rcl}
U&=&-e^{-u/\sqrt{\mu}}\\
V&=&-e^{v/\sqrt{\mu}},
\end{array}
\right.
\end{equation}  
where 
\begin{eqnarray}
\label{eq:kruskal1}
\left\{
\begin{array}{rcl}
u&=&t-r^*\\
v&=&t+r^*
\end{array}
\right.\\
r^*=r+\frac{\sqrt{\mu}}{2}\ln\left|\frac{\sqrt{\mu}-r}{\sqrt{\mu}+r}
\right|. \nonumber
\end{eqnarray}
In these coordinates the metric (\ref{eq:metric1}) takes the form
\begin{equation}
ds^2=-\frac{\mu (r+\sqrt{\mu})^2}{r^2 e^{2r/\sqrt{\mu}}} dUdV-r^2\left(
d\chi^2+\sin^2\chi(d\theta^2+\sin^2\theta d\phi^2 )\right).
\end{equation}
The line $l$ is defined by the equation $U-V=0$. The coordinate frame 
\begin{equation}
\label{eq:tild}
\left\{
\begin{array}{rcl}
T&=& \frac{\displaystyle 
e^{-R_l(\tau_0)/\sqrt{\mu}}\sqrt{\mu}
(R_l(\tau_0)+\sqrt{\mu})}
{\displaystyle 2R_l(\tau_0)}(U+V)
\\
X&=&\frac{\displaystyle 
e^{-R_l(\tau_0)/\sqrt{\mu}}\sqrt{\mu}
(R_l(\tau_0)+\sqrt{\mu})}
{\displaystyle 2R_l(\tau_0)} (U-V)
\end{array}
\right.
\end{equation}
is a comoving locally Minkowsky frame at the moment of time $\tau_0$.
If the trajectory of the surface
$M_{3+1}$ in this locally Minkowsky frame is  $(T(\tau), X(\tau))$,  
the size of the extra dimension is
\begin{equation}
\label{eq:size}
L(\tau_0)=2 X(\tau_0).
\end{equation}
The  coordinate $\tilde X=(U-V)/2$ changes along the brane trajectory 
according to the equation
\begin{equation}
\label{eq:tmp2}
\frac{d\tilde X}{d\tau}=\frac{1}{2\sqrt{\mu}}\left(
(U(\tau)-V(\tau))\frac{dr^*}{d\tau}-(U(\tau)+V(\tau))\frac{dt}{d\tau}\right)
\end{equation}
(see (\ref{eq:kruskal}), (\ref{eq:kruskal1})).
From  (\ref{eq:adot}) we find that in
first approximation in the  small parameter $m/M$
\begin{equation}
\label{eq:last1}
\frac{d \tilde X}{d\tau}+
\frac{R_l}{2\sqrt{\mu}(\mu-R_l^2)^{1/2}}\tilde X=
\frac{ G_5 m e^{R_l/\sqrt{\mu}}}
{\sqrt{\mu}(\sqrt{\mu}+R_l)^{3/2}(\sqrt{\mu}-R_l)^{1/2}},
\end{equation}
where $R_l$ is defined in (\ref{eq:zero}).
The last equation can be integrated 
\begin{equation}
\tilde X(R_l)=\frac{ G_5 m(\sqrt{\mu}-\sqrt{\mu-R^2_l})e^{R_l/\sqrt{\mu}}}
{\mu(\sqrt{\mu}+R_l)}
\end{equation}
and from (\ref{eq:tild}), (\ref{eq:size}) we find that the size $L$ of 
the extra dimension changes as 
\begin{equation}
\label{eq:thesize}
L=\frac{2 G_5 m(\sqrt{\mu}-\sqrt{\mu-R_l^2})}{R_l\sqrt{\mu}}
\end{equation}
It is equal to zero at the moment of the big bang $\tau=0$, 
then grows up to the maximal
value (\ref{eq:max}) at the moment of maximal expansion 
and then contracts back to zero at the moment of the big crunch. 

According to (\ref{eq:gg}) 
the  variation of the size of the extra dimension leads to a  variation of
the effective 4-dimensional gravitational constant in time
\begin{equation}
\label{eq:g3}
G_4\sim\frac{R_l\sqrt{\mu}}{m(\sqrt{\mu}-\sqrt{
\mu-R_l^2})}
\end{equation} 
It decreases when the universe expands.  

The analogous analysis can be done in the cases of  flat and open universes. 
In zero approximation (when the size of the extra dimension is equal to zero) 
the surface $M_{3+1}= M'_{3+1}$ follows the line $l:\{t=0\}$ of Fig. 
\ref{fig:brane3}.
 The scale factor changes with time as
\begin{equation}
\label{rl1}
R_l=\left\{
\begin{array}{ll}
\sqrt{2\sqrt{\mu}\tau}, & k=0\\
\sqrt{2\sqrt{\mu}\tau+\tau^2}, &k=-1.
\end{array}
\right.
\end{equation}
When the parameter $m/M$ is not 
equal to zero, but the energy of self-interaction of the universe 
through the compact extra dimension still dominates over all other kinds of
energy located at the surface $M_{3+1}$,  
the size of the extra dimension depends on the scale factor of the universe 
as
\begin{equation}
L(R_l)=\frac{2 G_5 m }{\sqrt{\mu}}
\left\{
\begin{array}{ll}
R_l/(2\sqrt{\mu}), &k=0\\ 
(\sqrt{\mu+R_l^2}-\sqrt{\mu})/R_l,& k=-1,
\end{array}
\right.
\end{equation}
it grows with time. But, in the case of an open universe it 
reaches asymptotically the finite value
\begin{equation}
L_f=\frac{2 G_5m}{\sqrt{\mu}},
\end{equation}
while in the case of a flat universe it grows unboundly.

To summarize, the gravitational self-interaction of the brane 
universe through the 
additional compact dimensions must be a general feature of higher-dimensional
cosmological models. This strong self-interaction could explain why 
the size of the extra dimensions is very small compared to the size of the 
(visible part of the) universe. The ``energy density'' 
(\ref{eq:effective}) of this self-interaction gives 
an essential contribution to the Friedmann equation (\ref{eq:adot}) which 
governs the dynamics of the scale factor of the universe.

\section{Acknowledgements.}

I am grateful to C.Armendariz Picon, V.F.Mukhanov, A.M.Boyarsky and 
D.V.Semikoz for the useful discussions on the 
subject. This work was supported by the
Sonderforschungsbereich SFB 375 f\"ur Astroteilchenphysik der Deutschen
Forschungsgemeinschaft.

\section{Appendix.}

In this Appendix we find solutions of the vacuum Einstein equations in 
$(4+1)$ dimensions. If we take the space-time metric in the form 
(\ref{eq:metric1}) the components of the Ricci tensor are
\begin{eqnarray}
R_{00}=\frac{1}{4}(\dot\lambda\dot\nu -\dot\lambda^2-2\ddot\lambda)
+\frac{1}{4}(\nu'^2-\nu'\lambda'+2\nu'')e^{\nu-\lambda}+\frac{3\nu'}{2r}
e^{\nu-\lambda}\nonumber\\
R_{11}=\frac{1}{4}(\lambda'\nu' -\nu'^2-2\nu'')
+\frac{1}{4}(\dot\lambda^2-\dot\nu\dot\lambda+2\ddot\lambda)e^{\lambda-\nu}
+\frac{3\lambda'}{2r}\nonumber\\
R_{01}=\frac{3\dot\lambda}{2r};\ \ \ 
R_{22}=\frac{1}{2}re^{-\lambda}(\lambda' -\nu')-2e^{-\lambda}
+2k\nonumber\\
R_{33}=fR_{22};\ \ \ R_{44}=f\sin^2\theta R_{22}\nonumber
\end{eqnarray}
where $k$ is defined right after equation (\ref{eq:solution}).
The relevant components of the Einstein equations which define the functions 
$\lambda,\nu$ in (\ref{eq:metric1}) are  
\begin{eqnarray}
\label{eins}
\left(\frac{\lambda'}{2r}-\frac{1}{r^2}\right) e^{-\lambda}+ \frac{k}{r^2}
={\cal T}_0^0\\
-\left(\frac{\nu'}{2r}+\frac{1}{r^2}\right) e^{-\lambda}+ \frac{k}{r^2}
={\cal T}_1^1.
\end{eqnarray}
Taking ${\cal T}_\alpha^\beta =0$ one  finds that for the vacuum solutions of
 Einstein equations 
\begin{equation}
e^\nu= e^{-\lambda}= k-\frac{\mu}{r^2}.
\end{equation}

\end{document}